\documentclass{PoS}

\usepackage{latexsym}
\usepackage{bm}
\usepackage{amssymb}
\usepackage{graphicx}

\newcommand{\case}[2]{\ensuremath{{\textstyle\frac{#1}{#2}}}}
\newcommand{\Dslash}{\ensuremath{{D\kern -0.65em /}}}

\newcommand{\half}{\ensuremath{{\textstyle\frac{1}{2}}}}

\newcommand{\quarter}{\ensuremath{{\textstyle\frac{1}{4}}}}
\newcommand{\sixth}{\ensuremath{{\textstyle\frac{1}{6}}}}

\newcommand{\ceight}{\ensuremath{r_E}}
\newcommand{\csix}{\ensuremath{z_E}}
\newcommand{\cone}{\ensuremath{c_2}}
\newcommand{\ctwo}{\ensuremath{c_1}}
\newcommand{\cfive}{\ensuremath{c_3}}
\newcommand{\cseven}{\ensuremath{z_3}}
\newcommand{\cthree}{\ensuremath{z_6}}
\newcommand{\cfour}{\ensuremath{c_4}}
\newcommand{\cnine}{\ensuremath{z_7}}
\newcommand{\cten}{\ensuremath{c_5}}
\newcommand{\celeven}{\ensuremath{r_5}}
\newcommand{\ctwelve}{\ensuremath{r_7}}
\newcommand{\cthirteen}{\ensuremath{r'_7}}

\newcommand{\cfifteen}{\ensuremath{z'_7}}
\newcommand{\rBB}{\ensuremath{r_{BB}}}
\newcommand{\zBB}{\ensuremath{z_{BB}}}
\newcommand{\cseventeen}{\ensuremath{c_{EE}}}
\newcommand{\rEE}{\ensuremath{r_{EE}}}
\newcommand{\zEE}{\ensuremath{z_{EE}}}

\title{An Improved Action for Heavy Quarks}

\ShortTitle{An Improved Action for Heavy Quarks}

\author{\speaker{Andreas S. Kronfeld} \\
	Theoretical Physics Department, 
	Fermi National Accelerator Laboratory,
	Batavia, Illinois, USA \\
	E-mail: \email{ask AT fnal.gov}}
\author{Mehmet B. Oktay \\
	School of Mathematics, Trinity College,
	Dublin, Ireland \\
	E-mail: \email{oktay AT maths.tcd.ie}}

\abstract{We extend the Fermilab method for heavy quarks to include all 
interactions of dimension six in the action.
We discuss a subtlety in the power counting, which implies that, for
heavy quarks, certain interactions of dimension seven are commensurate
with some of those of dimension six.
We then present tree-level matching conditions obtained from 
calculating the Compton scattering amplitude for (lattice) QCD.
When the matching conditions have been applied, the improved action
removes (tree-level) discretization errors of order~$a^2\bm{p}^3/m_Q$
and~$a^3\bm{p}^3$.}

\FullConference{XXIVth International Symposium on Lattice Field Theory\\
                July 23--28, 2006\\
                Tucson, Arizona, USA}

\begin{document}

\section{Introduction}

Discretization errors are an important source of uncertainty in lattice
QCD calculations with heavy quarks.
As with gluons and light quarks, one uses effective field theory to
understand and control discretization (aka cutoff or lattice-spacing)
effects.
With heavy quarks there are two possibilities: a (modified) Symanzik
effective Lagrangian and a (modified) heavy-quark effective Lagrangian
(HQET for heavy-light hadrons; NRQCD for quarkonium).

Usually, the Symanzik local effective Lagrangian (LE$\mathcal{L}$) is 
devised with only one short distance in mind, the lattice spacing $a$.
Similarly, HQET/NRQCD is usually constructed with only the heavy quark's
(or quarks') Compton wavelength(s), $1/m_Q$, as short distance(s).
With heavy quarks on a lattice one must consider both kinds of short
distances.
In particular, couplings in the lattice Lagrangian and Wilson
coefficients in the effective Lagrangians depend on the dimensionless
ratio(s)~$m_Qa$.

Treating the inverse quark mass as a short distance, the Symanzik
LE$\mathcal{L}$ can be written~\cite{Kronfeld:2002pi}
\begin{eqnarray}
	\mathcal{L}_{\rm Sym} & = & - \bar{q}\left(\gamma_4D_4 +
		\sqrt{\frac{m_1}{m_2}} \bm{\gamma}\cdot\bm{D} + m_1 \right)q +
		aK_t^{\rm lat}\bar{q}(\gamma_4D_4 + m_1)^2q +
		aK_s^{\rm lat}\bar{q}\bm{D}^2q \nonumber \\
	& + &
		aK_B^{\rm lat}\bar{q}i\bm{\Sigma}\cdot\bm{B}q +
		aK_E^{\rm lat}\bar{q} \bm{\alpha}\cdot\bm{E}q + \cdots,
	\label{eq:Sym}
\end{eqnarray}
where $m_1$ is the renormalized rest mass, and mass-dependent 
short-distance effects are lumped into $\sqrt{m_1/m_2}$ and the 
coefficients $K^{\rm lat}$.
An improved action is devised by adjusting $\mathcal{L}_{\rm Sym}$ to 
reproduce QCD.
Some coefficients may be made to vanish via field redefinitions, 
such as~$K_t^{\rm lat}$ and 
$K_s^{\rm lat}$~\cite{Sheikholeslami:1985ij,El-Khadra:1996mp}.
The others must be addressed by matching the underlying lattice 
Lagrangian:
\begin{eqnarray}
	m_1 = m_2 = m_Q, \quad
	K_B^{\rm lat}(c_B) & = & 0, \quad K_E^{\rm lat}(c_E) = 0, \ldots,
	\label{eq:K0}
\end{eqnarray}
where $m_Q$ is the physical quark mass, and $c_B$ and $c_E$ are
couplings of dimension-five interactions in the Fermilab lattice action
for heavy quarks~\cite{El-Khadra:1996mp}.

Treating the lattice spacing and inverse heavy quark mass as short 
distances, the heavy-quark effective Lagrangian is~\cite{Kronfeld:2000ck}
\begin{eqnarray}
	\mathcal{L}_{\rm HQ} & = & - \bar{h}^{(\pm)}\left(\gamma_4D_4 +
		m_1 \right)h^{(\pm)} +
		\frac{\bar{h}^{(\pm)}\bm{D}^2h^{(\pm)}}{2m_2} +
		\frac{Z_B^{\rm lat}\bar{h}^{(\pm)}i\bm{\Sigma}\cdot\bm{B}h^{(\pm)}}{2m_2} \nonumber \\
	& + &
		\frac{Z_E^{\rm lat}\bar{h}^{(\pm)}i\bm{\Sigma}\cdot[
			\bm{D}\times\bm{E}]h^{(\pm)}}{8m_2^2} + 
		\frac{Z_D^{\rm lat}\bar{h}^{(\pm)}
			\bm{D}\cdot\bm{E}h^{(\pm)}}{8m_2^2} + \cdots,
	\label{eq:HQ}
\end{eqnarray}
where the $Z^{\rm lat}$ capture the short-distance effects.
In parallel, one can construct this effective Lagrangian for continuum
QCD, with the same operators but different coefficients~$Z^{\rm cont}$.
Now improvement is achieved by choosing improvement couplings in the
underlying lattice Lagrangian, so that the two effective HQ Lagrangians
coincide.
That is
\begin{eqnarray}
	m_2 = m_Q, \quad Z_B^{\rm lat}(c_B) & = & Z_B^{\rm cont}, \quad 
		Z_E^{\rm lat}(c_E) = Z_E^{\rm cont}, \ldots.
	\label{eq:ZZ}
\end{eqnarray}
Both sets of conditions, Eqs.~(\ref{eq:K0}) or Eqs.~(\ref{eq:ZZ}), yield
the same results for the improvement couplings $c_B$ and $c_E$.

In this paper we focus on an extension of the Fermilab method to 
include all interactions of dimension~six.
Some aspects have been reported earlier~\cite{Oktay:2002mj,Oktay:2003gk}.
Here, we first discuss a subtlety in the power counting, which implies that 
we must also consider certain interactions of dimension seven to be 
commensurate with some of those of dimension six.

\section{Power Counting}

With heavy quarks one must pay special attention to power counting.
This is simplest in heavy-light hadrons, where the typical
three-momentum in the rest frame is $p\approx\Lambda_{\rm QCD}$.
Then it is useful to define $n_\Gamma$, which is 0 or 1, depending on
whether the Dirac matrix~$\Gamma$ in a quark bilinear commutes or
anticommutes with~$\gamma_4$.
A bilinear of dimension $d$ introduces physical effects of order
\begin{equation}
	(a\Lambda_{\rm QCD})^{d-4}
	\left(\frac{\Lambda_{\rm QCD}}{m_Q}\right)^{n_\Gamma} \sim
	(a\Lambda_{\rm QCD})^{d-4+n_\Gamma} \sim
	\left(\frac{\Lambda_{\rm QCD}}{m_Q}\right)^{d-4+n_\Gamma} ,
	\label{eq:power}
\end{equation}
which are all the same once one allows for coefficient functions that
depend on $m_Qa$, with $m_Qa$ of order unity.
This power counting has recently been considered by Christ, Li, and
Lin~\cite{Christ:2006us}.
It is useful for hadrons with one heavy quark, but for quarkonium one
should adopt the power counting of nonrelativistic
QCD~\cite{Lepage:1992tx}.

We shall denote terms in the lattice action by $S_{(d,n_\Gamma)}$ to
classify them by their power counting~(\ref{eq:power}).
The terms with $n_\Gamma=1$ are necessary to ensure a smooth limit when
$m_Qa\to0$~\cite{El-Khadra:1996mp}, which is a feature distinguishing
the Fermilab method from lattice NRQCD.
Ref.~\cite{El-Khadra:1996mp} treated $S_{(5,0)}$ and $S_{(5,1)}$ 
(and cursorily $S_{(6,0)}$).
Here we complete the analysis presented at earlier Lattice
symposia~\cite{Oktay:2002mj,Oktay:2003gk} to encompass the full set of
interactions in $S_{(6,1)}$ and $S_{(7,0)}$.

\section{Improved Fermilab Action}

The Fermilab action is a generalization of the Sheikholeslami-Wohlert
action.
We write
\begin{equation}
	S = S_0 + \sum_{d=5}^{\infty}\sum_{n_\Gamma=0}^1 S_{(d,n_\Gamma)},
	\label{eq:S}
\end{equation}
where
\begin{eqnarray}
	S_0 & = & m_0a^4\sum_x\bar{\psi}(x)\psi(x) +
		a^4\sum_x\bar{\psi}(x) \gamma_4{D_4}_{\mathrm{lat}} \psi(x) -
		\half a^5\sum_x 
			\bar{\psi}(x){\triangle_4}_{\mathrm{lat}}\psi(x) 
		\nonumber \\ & & +\,
		\zeta a^4\sum_x
			\bar{\psi}(x)\bm{\gamma}\cdot\bm{D}_{\mathrm{lat}} \psi(x) -
		\half r_s\zeta a^5\sum_x 
			\bar{\psi}(x)\triangle^{(3)}_{\mathrm{lat}}\psi(x) .
	\label{eq:S0} 
\end{eqnarray}
We denote lattice fermions fields with $\psi$ to distinguish them from
the continuum quark fields in Eqs.~(\ref{eq:Sym}) and~(\ref{eq:HQ}).
The coupling $\zeta$ allows one to adjust $m_2=m_1$ in Eq.~(\ref{eq:Sym}).
The notation for $\bm{D}_{\rm lat}$, $\triangle^{(3)}_{\rm lat}$, etc.,
is as in Ref.~\cite{El-Khadra:1996mp}.
The dimension-five interactions 
are~\cite{Sheikholeslami:1985ij,El-Khadra:1996mp}
\begin{eqnarray}
	S_{(5,0)} = S_B & = & -\half c_B\zeta a^5 \sum_x 
		\bar{\psi}(x) i\bm{\Sigma}\cdot\bm{B}_{\mathrm{lat}} \psi(x),
	\label{eq:SB} \\ 
	S_{(5,1)} = S_E & = & -\half c_E\zeta a^5 \sum_x 
		\bar{\psi}(x)  \bm{\alpha}\cdot\bm{E}_{\mathrm{lat}} \psi(x),
	\label{eq:SE}
\end{eqnarray}
where the notation $S_B$ and $S_E$ is from Ref.~\cite{El-Khadra:1996mp}.
At dimension six and seven we introduce
\begin{eqnarray}
	S_{(6,0)} & = &
		\ceight a^6 \sum_x \bar{\psi}(x) 
			\{\bm{\gamma}\cdot \bm{D}_{\mathrm{lat}},
			\bm{\alpha}\cdot \bm{E}_{\mathrm{lat}}\}
			\psi(x) \nonumber \\
	& + &
		\csix a^6 \sum_x \bar{\psi}(x)\gamma_4 \left(
			\bm{D}_{\mathrm{lat}}\cdot\bm{E}_{\mathrm{lat}} -
			\bm{E}_{\mathrm{lat}}\cdot\bm{D}_{\mathrm{lat}}
			\right) \psi(x), \label{eq:S60} \\ 
	S_{(6,1)} & = & 
		\ctwo a^6 \sum_x \bar{\psi}(x) 
			\sum_i\gamma_i {D_i}_{\mathrm{lat}}
			{\triangle_i}_{\mathrm{lat}} \psi(x) \nonumber \\
	& + & 
		\cone a^6 \sum_x \bar{\psi}(x) 
			\{\bm{\gamma}\cdot \bm{D}_{\mathrm{lat}},
			\triangle^{(3)}_{\mathrm{lat}}\} \psi(x) \nonumber \\
	& + &
		\cfive a^6 \sum_x \bar{\psi}(x) 
			\{\bm{\gamma}\cdot \bm{D}_{\mathrm{lat}},
			 i\bm{\Sigma}\cdot \bm{B}_{\mathrm{lat}}\}
			 \psi(x) \nonumber \\ 
	 & + &
		 \cseven a^6 \sum_x \bar{\psi}(x) \bm{\gamma}\cdot \left(
			 \bm{D}_{\mathrm{lat}}\times\bm{B}_{\mathrm{lat}} +
			 \bm{B}_{\mathrm{lat}}\times\bm{D}_{\mathrm{lat}}
			 \right) \psi(x) \nonumber \\
	 & + &
		 \cseventeen a^6 \sum_x \bar{\psi}(x) 
			 \{ \gamma_4{D_4}_{\mathrm{lat}},
			 \bm{\alpha}\cdot\bm{E}_{\mathrm{lat}}\}
			 \psi(x), \label{eq:S61} \\
	S_{(7,0)} & = & 
		\cfour a^7 \sum_x \bar{\psi}(x) 
			\sum_i {\triangle_i}_{\mathrm{lat}}^2
			\psi(x) \nonumber \\
	& + &
		\cten a^7 \sum_x \bar{\psi}(x) \sum_i \sum_{j\neq i}
			\{ i\Sigma_i {B_i}_{\mathrm{lat}},
				{\triangle_j}_{\mathrm{lat}} \} 
			\psi(x)\nonumber  \\
	& + &
		\celeven a^7 \sum_x \bar{\psi}(x) \sum_i \sum_{j\neq i}
			i\Sigma_i \left[D_j B_i	D_j\right]_{\mathrm{lat}}
			\psi(x) \nonumber \\
	& + &
		\cthree a^7 \sum_x \bar{\psi}(x) \left( 
			\triangle^{(3)}_{\mathrm{lat}}\right)^2 
			\psi(x) \nonumber \\
	& + &
		\cnine a^7 \sum_x \bar{\psi}(x) 
			\{\triangle^{(3)}_{\mathrm{lat}},
			 i\bm{\Sigma}\cdot \bm{B}_{\mathrm{lat}}\} 
			\psi(x) \nonumber \\
	& + &
		\cfifteen a^7 \sum_x \bar{\psi}(x) [D_i
			i\bm{\Sigma}\cdot\bm{B}D_i]_{\mathrm{lat}}
			\psi(x) \nonumber \\
	& + & 
		\ctwelve a^7 \sum_x \bar{\psi}(x) 
			\bm{\gamma}\cdot\bm{D}_{\mathrm{lat}}
			i\bm{\Sigma}\cdot\bm{B}_{\mathrm{lat}}
			\bm{\gamma}\cdot\bm{D}_{\mathrm{lat}}
			\psi(x) \nonumber \\
	& + &
		\cthirteen a^7 \sum_x \bar{\psi}(x) [\bm{D}\cdot\left(
			\bm{B}\times\bm{D}\right)]_{\mathrm{lat}} 
			\psi(x) \nonumber \\
	& + &
		\rBB a^7 \sum_x \bar{\psi}(x) \left(
			i\bm{\Sigma}\cdot\bm{B}_{\mathrm{lat}} \right)^2
			\psi(x) \nonumber \\
	& + &
		\zBB a^7 \sum_x \bar{\psi}(x) 
			\bm{B}_{\mathrm{lat}}\cdot\bm{B}_{\mathrm{lat}} 
			\psi(x) \nonumber \\
	& - &
		\rEE a^7 \sum_x \bar{\psi}(x) \left(
			\bm{\alpha}\cdot\bm{E}_{\mathrm{lat}} \right)^2 
			\psi(x) \nonumber \\
	& + &
		\zEE a^7 \sum_x \bar{\psi}(x) 
			\bm{E}_{\mathrm{lat}}\cdot\bm{E}_{\mathrm{lat}} 
			\psi(x). \label{eq:S70}
\end{eqnarray}
In particular, in heavy-quark power counting the interactions in
$S_{(7,0)}$ are commensurate with those in~$S_{(6,1)}$.
Most of these interactions have been discussed
previously~\cite{Oktay:2002mj,Oktay:2003gk}, but the following is the
first discussion of interactions with two chromomagnetic or
chromoelectric fields, and the related interaction with
coupling~$\cseventeen$.%
\footnote{At dimension six and higher one must also consider four-quark
interactions, which we shall discuss elsewhere.}

By applying arbitrary field redefinitions one can make certain
coefficients in the effective Lagrangian vanish, and their interactions
are called \emph{redundant}.
A corresponding number of couplings in the lattice action may be
chosen by convenience.
For example, to obviate the doubling problem one sets $r_s\neq0$
and~$r_t=1$ in Eq.~(\ref{eq:S0}).
One can show that \emph{all} interactions with further time derivatives
acting on quark and antiquark fields are redundant, so we omit them from
Eqs.~(\ref{eq:SB})--(\ref{eq:S70}).
Among the rest there is some freedom on which ones to call redundant; 
with an eye towards efficiency in computing quark propagators we choose 
\begin{equation}
	\ceight = \celeven = \ctwelve = \cthirteen = \rBB = \rEE = 0,
\end{equation}
a choice that is substantiated further by the matching calculations 
discussed in the next section.

\section{Matching}

To find out how to adjust the other couplings, we must explicitly derive
Eqs.~(\ref{eq:K0}) or, equivalently, Eqs.~(\ref{eq:ZZ}), and solve them 
for the couplings in the lattice action.
Because the action includes interactions with two chromoelectric and two
chromomagnetic fields, we have done so by working out the amplitude for Compton
scattering.
We find, not unexpectedly, that many couplings vanish at the tree level.
To be specific, tree-level matching yields
\begin{equation}
	\csix = \cseven = \cthree = \cnine = \cfifteen = \zBB = \zEE = 0,
\end{equation}
but these become non-trivial beyond the tree level.

The other couplings---$c_B$, $c_E$, $c_{EE}$, $c_1$, $c_2$, $c_3$,
$c_4$, and~$c_5$---must be non-zero, with an explicit, sometimes nontrivial,
dependence on $(m_0a,\zeta,r_s)$.
We do not have space to discuss the intermediate steps here, so we 
quote the final result of tree-level matching.
We find
\begin{eqnarray}
	c_B & = & r_s, \\
	c_E & = & \frac{\zeta^2-1}{m_0a(2+m_0a)} + 
		\frac{r_s\zeta}{1+m_0a} + 
		\frac{r_s^2m_0a(2+m_0a)}{4(1+m_0a)^2}  , \\
	\ctwo & = & - \sixth \zeta +
		c_B \frac{m_0a(2+m_0a)}{6(1+m_0a)}, \\
	\cone = \cfive & = & 
		\frac{\zeta^3(\zeta^2-1)}{[2m_0a(2+m_0a)]^2} - 
		\frac{\zeta^2[\zeta+2r_s(1+m_0a)-3r_s\zeta^2/(1+m_0a)]}%
		{8m_0a(2+m_0a)} \nonumber \\
	& + & \frac{3r_s^2\zeta^3}{16(1+m_0a)^2} +
		\frac{m_0a(2+m_0a)r_s^2\zeta}{32(1+m_0a)^2}\left[
		\frac{r_s\zeta}{1+m_0a} - 1 \right], \\
	\cseventeen [2+m_0a(2+m_0a)] & = & 
		\frac{\zeta(\zeta^2-1)(1+m_0a)}{[m_0a(2+m_0a)]^2} +
		\frac{c_E\zeta(\zeta^2-1)(1+m_0a)}{m_0a(2+m_0a)} \nonumber \\
		& + & \frac{\zeta (r_s\zeta - 1 - m_0a)}{2m_0a(2+m_0a)} + 
			\half r_sc_E\zeta^2 - \quarter c_E^2\zeta(1+m_0a) , \\
	\cfour & = & \case{1}{24} r_s\zeta + \case{1}{3} c_B\zeta , \\
	\cten & = & \case{1}{4}c_B\zeta . 
\end{eqnarray}
Some aspects are not surprising: for example $\cone=\cfive$, which makes 
sense because together they provide an interaction of the form
$\bar{\psi}(\bm{\gamma}\cdot\bm{D})^3\psi$.
Two out of the three couplings $(\cfour,\cten,\celeven)$ must be 
nonzero, essentially to correct errors from the $\bm{B}_{\rm lat}$ in 
$S_{(5,0)}$.
We choose $\cfour\neq0$ because we expect its interaction to be easier 
to compute than that with coupling~$\celeven$. 

\section{Outlook}

In several recent calculations with Fermilab quarks, the largest 
systematic uncertainty comes from a (conservative) estimate of 
heavy-quark discretization effects.
With the improved action presented here, the same technique for
estimating the uncertainties~\cite{Kronfeld:2003sd} suggests that these
effects are reduced to a few percent~\cite{Oktay:2003gk}.
To achieve this target, tree-level matching should suffice for 
$S_{(6,1)}$ and $S_{(7,0)}$.
For $S_{(5,0)}$, $S_{(5,1)}$, and $S_{(6,0)}$ higher accuracy is needed,
either at the one-loop~\cite{Nobes:2003nc,Aoki:2003dg} or
(for $c_B$ in $S_{(5,0)}$) nonperturbative level~\cite{Lin:2006ur}.

\section*{Acknowledgments}

M.B.O. is supported by Science Foundation Ireland grant~04/BRG/P0275.
Fermilab is operated by Universities Research Association Inc., under
contract with the U.S.\ Department of Energy.

\end{document}